\documentstyle[emulateapj,pstricks]{article}

\def\ev{\mbox{eV}}
\def\fsl{\mbox{$R_{f,WDM}$}}
\def\hubble{\mbox{km s$^{-1}$ Mpc$^{-1}$}}
\def\kev{\mbox{keV}}
\def\kmpch{\mbox{h Mpc$^{-1}$}}
\def\kms{\mbox{km/s}}
\def\kpch{\mbox{$h^{-1}$kpc}}
\def\mpc{\mbox{Mpc}}
\def\mpch{\mbox{$h^{-1}$Mpc}}
\def\msun{\mbox{M$_\odot$}}	
\def\msunh{\mbox{$h^{-1}$M$_\odot$}}
\def\mev{\mbox{MeV}}
\def\ome{\mbox{$\Omega_0$}}
\def\omel{\mbox{$\Omega_\Lambda$}}
\def\omew{\mbox{$\Omega_{WDM}$}}
\def\sige{\mbox{$\sigma_8$}}
\def\vmax{\mbox{$V_{\rm max}$}}
\def\mathnew{\mathsurround=0pt}
\def\ref{\par\noindent\hangindent=2pc \hangafter=1 }
\def\simov#1#2{\lower .5pt\vbox{\baselineskip0pt
    \lineskip-.5pt\ialign{$\mathnew#1\hfil##\hfil$\crcr#2\crcr\sim\crcr}}}  
\def\simgreat{\mathrel{\mathpalette\simov >}}
\def\simless{\mathrel{\mathpalette\simov <}}
\def\'#1{\ifx#1i{\accent"13\i}\else{\accent"13#1}\fi}
\def\eg{e.g.,}

\def\ie{i.e.}
\begin{document}
\slugcomment{{\em Astrophysical Journal, accepted}}
\lefthead{SUBSTRUCTURE IN A WARM DARK MATTER COSMOLOGY}
\righthead{COLIN, AVILA-REESE, \& VALENZUELA}
 
\title{Substructure and halo density profiles in a Warm Dark Matter
Cosmology}

\author{Pedro Col\'in\altaffilmark{1} and Vladimir Avila-Reese\altaffilmark{2}}
\affil{Instituto de Astronom\'ia, Universidad Nacional Aut\'onoma
de M\'exico, C.P. 04510, M\'exico, D.F., M\'exico}

\author{and} 

\author{Octavio Valenzuela\altaffilmark{3}}
\affil{Astronomy Department, New Mexico State University, Box 30001, Department
4500, Las Cruces, NM 88003-0001}

\altaffiltext{1}{colin@astroscu.unam.mx}
\altaffiltext{2}{avila@astroscu.unam.mx}
\altaffiltext{3}{ovalenzu@nmsu.edu}

\begin{abstract}
       We performed a series of high-resolution collisionless N-body simulations designed 
to study the substructure of Milky Way-size galactic halos (host halos) 
and the density profiles of halos in a warm dark matter (WDM) scenario 
with a non-vanishing cosmological constant. The virial masses of the host halos 
range from $3.5 \times 10^{12}\ \msunh$ to $1.7 \times 10^{12}\ \msunh$ 
and they have more than $10^5$ particles each. A key feature of the 
WDM power spectrum is the free-streaming length \fsl\ which fixes an 
additional parameter for the model of structure formation. We analyze 
the substructure of host halos using three \fsl\ values:  0.2, 0.1, 
and 0.05 \mpc\ and compare results
to the predictions of the cold dark matter (CDM) model. 
We find that guest halos (satellites) do form in the WDM scenario but
are more easily destroyed by dynamical friction
and tidal disruption than their counterparts in a CDM model. The small 
number of guest halos that we find in the WDM models with respect to 
the CDM one is the result of a lower guest halo accretion and a higher 
satellite destruction rate.
These two phenomena operate almost with the same intensity in delivering 
a reduced number of guest halos at $z = 0$. For the model with 
$\fsl = 0.1\ \mpc$ the number of accreted small halos is a factor 2.5 
below that of the CDM model while the fraction of destroyed satellites
is almost twice larger than that of the CDM model.
The larger the \fsl\ value the greater the size of these two effects 
and the smaller the abundance of satellites.
Under the assumption that each guest halo hosts a luminous galaxy, we 
find that the observed circular velocity function of satellites 
around the Milky Way 
and Andromeda is well described by the $\fsl = 0.1\ \mpc$ WDM model. 
In the $\fsl =0.1- 0.2\ \mpc$ models, the surviving guest halos at $z=0$
---whose masses are in the range  $M_h \approx 10^9-10^{11}\ \msunh$--- have an
average concentration parameter $c_{1/5}$ $( =r(M_h)/r(M_h/5) )$ 
which is approximately twice smaller than that of the corresponding
CDM guest halos. This difference, very likely, produces the higher 
satellite destruction rate found in the WDM models.
The density profile of host halos is well described by the NFW fit
whereas guest halos show a wide variety of density profiles. 
A tendency to form shallow cores is not evident; the
profiles, however, are limited by a poor mass resolution in the innermost
regions were shallow cores could be expected. 
\end{abstract}

%=====================

\section{Introduction}

%=====================

Non-baryonic dark matter is an essential ingredient of current
inflation-inspired models of cosmic structure formation 
in the universe. 
From the point of view of particle physics,
there is no obvious preference for any of the
predicted dark matter candidates (\cite{CDW96}), which, according to 
their rms velocity at the time of their decoupling,
can be cold, warm, or hot. From the point of view of
structure formation, the most compelling candidate has been
the cold dark matter. 
The CDM scenario for structure formation has successfully 
accounted for several observational facts, particularly on large 
scales, without introducing an additional free parameter related 
to its particle distribution function in phase space.
However, on small scales and/or in high-density regions 
of the universe, the predictions of the CDM models seem to be in conflict 
with observations. 
  
One of the potential problems of the CDM scenario is
that the predicted number of low-mass halos ---where probably dwarf
galaxies form--- within a Milky Way-size halo, greatly exceeds 
the observed abundance of satellite galaxies in the Local Group 
(Klypin et al. 1999, hereafter \cite{KKVP99}; \cite{Moore99a}; 
see also \cite{Kauffmann93}).
A second problem is that the predicted inner density profiles of 
CDM halos may disagree with the shallow profiles inferred from 
the rotation curves of dwarf and low surface brightness galaxies 
(\cite{Moore94}; \cite{FP94}; \cite{Burkert95}; \cite{dBM97}; \cite{HG98}), 
although the observational data for the latter galaxies are 
controversial (\cite{vdB99}; \cite{SMT2000};
but see \cite{Firmani2000b}). High-resolution gravitational lensing 
maps of a cluster of galaxies have also revealed a soft inner mass
distribution in the halo of this cluster (\cite{TKD98}). 
The rotation curve decompositions of 
normal galaxies and the Tully-Fisher relation obtained in galaxy
formation models as well as the dark mass contained within 
the solar radius in our Galaxy, also point out to dark halos shallower 
and/or much less concentrated than those 
predicted by the CDM model (\cite{AFH98}; \cite{Navarro98}; \cite{NS99}; 
\cite{FA2000}; \cite{MM2000}). If these shortcomings 
of the CDM scenario are confirmed with more observational and theoretical 
data, new alternatives (cosmological and/or astrophysical) have to
be explored in order to modify the properties of the mass distribution
at small scales.  

In a recent burst of papers, explored alternatives include modifications
to: either the nature of the dark matter candidate (\eg\ \cite{SS99}; 
Hannestad 1999; \cite{SomDol00}; \cite{WC2000}; \cite{Firmani2000a}; 
\cite{HD2000}; \cite{Moore2000}; \cite{YSWT2000} ; \cite{Burkert2000}; 
\cite{Peebles2000}; Hannestad \& Scherrer 2000; Riotto \& Tkachev 2000),
or the generation of 
the primordial power spectrum (\eg\ \cite{KL99}). More 
conservative astrophysical mechanisms to overcome the
problems mentioned above have also been proposed (\eg\ \cite{NEF96}; \cite{GS99};
\cite{BKW2000}; \cite{BGS2000}). One possible modification is to 
go from a CDM scenario to a warm dark matter (WDM) one.
The WDM particles (warmons)
would suppress the power at small scales by free-streaming out
of overdense regions limiting the
formation of substructure at scales below the free-streaming scale. 
At large scales, the structure formation 
would proceed in a very similar way to that of a CDM model. N-body 
simulations have shown
that indeed large-scale structure in WDM models looks 
similar to that of a CDM model (Colombi et al. 1996). On the other
hand, as Hogan \& Dalcanton (2000) noted, the finite phase 
density of dark halos inferred from observations could be pointing 
to a non-negligible DM velocity dispersion at the time of structure
formation.

Using the Press-Schechter formalism, Kamionkowski \& Liddle 
(1999) have shown that if the CDM power spectrum is filtered at 
scales corresponding to dwarf galaxies, then the abundance
of Milky Way satellites can be reproduced. Recently, \cite{WC2000}
reported results from N-body simulations for WDM models at high
redshifts. They found that the abundance of $10^{10} \msunh$ halos
is reduced by a factor of $\sim 5$ at $z = 3$ with respect to the CDM model
when the power spectrum is filtered at $k \approx 2 \kmpch$. At the same time
they showed that the Ly-$\alpha$ power spectrum at this redshift is very 
similar to that of the CDM model, which is in agreement with 
observations. This apparent contradictory
result is explained by the fact that the collapse of large-scale structures,
as they go non-linear, regenerates the initially suppressed small-scale
modes in the power spectrum (\cite{WC2000}). 

These results encourage us
to explore in more detail the predictions of WDM N-body
simulations at the present epoch. 
Does the suppression of power at small scales of a WDM model
actually eliminate the excessive degree of substructure
predicted by the CDM scenario? 
Are the WDM halos less concentrated? And if so, do they have a smoother 
inner mass distribution than their counterpart CDM halos? 
The main aim of this paper is to give a quantitative answer
to the first question. To this end we have carried out high-resolution N-body
simulations of Milky Way-size galactic halos 
in three different WDM models. A host halo 
of about $2 \times 10^{12}\ \msunh$ has more
than $10^5$ particles in the simulations. 
Since the most successful variant of the CDM models 
is a flat universe with a non-zero cosmological constant 
($\Omega_{\Lambda} = 0.7$ and $h = 0.7$), here we also use 
this cosmological model 
but instead of CDM we introduce WDM with the extra free
parameter \fsl: these models will be our $\Lambda$WDM models (for
economy we drop off the greek letter $\Lambda$ hereafter when
we refer to either CDM or WDM models).
We will also address the questions of concentrations and 
density profiles of dark halos, although the small number of 
large high-resolved halos and the small range of masses in the simulations 
constrain our predictions on this subject.  

In Section 2 we discuss the WDM models to be explored
in this paper. In Section 3 we briefly describe the numerical technique that
we used for the simulations. Section 4 is devoted to the analysis 
and comparison with observations of the circular velocity function of 
satellites within host halos of Milky Way-sizes. The concentrations and
density profiles of the host and satellite halos are presented in 
Section 5. In Section 6 we discuss some of the results, and summarize of 
our main conclusions is given in Section 7.
 
%============================================

\section{The $\Lambda$WDM Cosmological Models}

%============================================

Several observational
tests such as the distribution of galaxies (\eg\ Peacock \& Dodds 1994),
cluster mass estimates (\eg\ \cite{Carlberg96}), the 
determination of the baryon
fraction in clusters (\eg\ \cite{MME99}), and the evolution of cluster
abundance (\eg\ \cite{Eke98}) point to a cosmological CDM model with
a low matter density, $\ome\ \approx 0.3$. This model also successfully 
accounts for the observationally inferred 
values of the Hubble constant and the age of the universe. On the other
hand, according to a prediction of the inflationary theory,
the universe should be flat, \ie\ a contribution to the
density of the universe from a cosmological 
constant is necessary if $\ome \approx 0.3$. It has been inferred
recently from observations of high redshift Supernovae  
that the universe is expanding with 
positive acceleration (\cite{Perlmutter99}; \cite{Riess98}; \cite{Schmidt98}).
Remarkably, the estimated value of the cosmological constant density parameter 
is in this case $\omel \approx 0.6-0.8$. Thus, the 
most popular cosmological model has become a flat CDM model with a non-vanishing
cosmological constant: $\ome \approx 0.3$, $\omel \approx 0.7$, and
$h = 0.7$ (the Hubble constant in units of $100\ \hubble$; this
value is consistent with the current observational determinations, 
\eg\ \cite{NR98}). Here we will use this cosmological model but 
instead of CDM we will introduce WDM. 

A dark matter particle is usually defined as {\it hot} or {\it cold} 
if at the moment of decoupling from the rest of the cosmic plasma it is
relativistic or non-relativistic, respectively (\cite{KolbTurner90}).
The classic and only example of detected dark matter are the neutrinos. 
They are hot because they were relativistic at
the moment of their decoupling. If the mass of the dark matter 
candidate is much higher than 1 GeV and the strength of its interactions
is comparable to that of the weak interaction, then it would behave as 
cold dark matter. The thermal velocities of these particles at the 
time of structure formation is negligible.  In this paper we are 
interested in a {\it warm} DM candidate, a thermal relic that 
at the time of its decoupling was relativistic and whose
mass $m_W$ is much higher than that of its hot counterpart.
The Cowsik-McClelland bound prohibits any candidate with a
mass larger than $\sim 15\ \ev$  (assuming $\omew \sim 0.3$ and $h= 0.7$) 
which decouples when the temperature of the universe was a few  \mev\
(\cite{CowMc72}). Thus, the warmon should decouple earlier
than a hot candidate, in an epoch when the total number of 
degrees of freedom of relativistic particles was certainly very high
(\cite{KolbTurner90}). 

Unlike the CDM case, the small-scale density fluctuations are damped 
out in a WDM scenario 
by the free-streaming of DM particles. It is straightforward to compute
the comoving free-streaming scale \fsl\ (\eg\ \cite{SomDol00}):
\begin{equation}
\fsl = 0.2\ (\omew h^2)^{1/3}\ \left( \frac{m_W}{1 \kev} \right)^{-4/3} \mpc.
\end{equation}
\noindent The WDM scenario was not attractive in the past 
because of the introduction
of the extra free parameter \fsl, and because particles in the 
required mass range of $\sim 100\ \ev -1\ \kev$ were not particularly compelling.  
Nevertheless, these arguments are somewhat obsolete nowadays. As mentioned
in the introduction, the CDM scenario seems to be in disagreement with
observational data at small scales, so that models with
extra degrees of freedom might be necessary. On the other hand,
light WDM candidates as palatable as the CDM ones are also predicted
by particle physics beyond the standard model; one possible
example are the right-handed neutrinos (\eg\ \cite{CDW96}).
\begin{figure*}[ht]
%\pspicture(0.5,-1.5)(15.0,19.0)
\pspicture(0.5,-1.5)(13.0,19.0)
%\rput[tl]{0}(1.0,18.0){\epsfxsize=18cm
\rput[tl]{0}(2.8,17.0){\epsfxsize=14cm
\epsffile{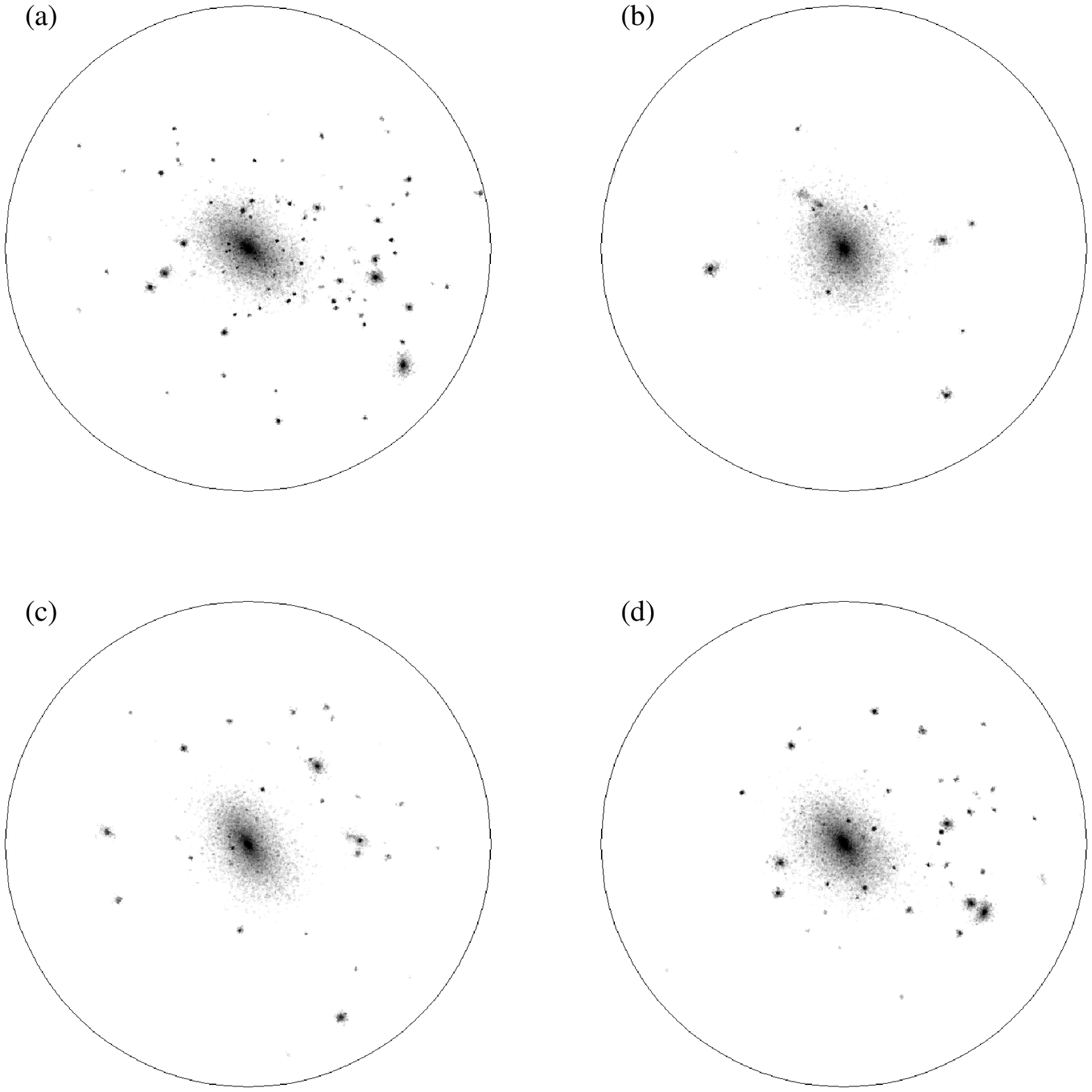}}
%\rput[tl]{0}(0.5,0.0){
\rput[tl]{0}(0.5,2.0){
%\begin{minipage}{18cm}
\begin{minipage}{18.4cm}
  \small\parindent=3.5mm {\sc Fig.}~1.---
The distribution of dark matter particles inside a
sphere of 400 \kpch\ radius (solid circle) for the host halo II in four
different models: (a) CDM model, and (b), (c), and (d) the WDM models with
$\fsl = 0.2$, 0.1, and 0.05 \mpc, respectively. To enhance the contrast
we have color-coded DM particles on a grey scale according to their local
density (a pgplot code kindly provided by A. Kravtsov) and removed all
DM particles whose density was lower than a certain value. The local 
density at the particle positions was computed using SMOOTH, a 
publicly available code developed by the HPCC group in the UW Department 
of Astronomy.
\end{minipage}}
\endpspicture
\end{figure*}

We have represented the WDM power spectrum by the following expression
(\cite{BBKS86}; see also \cite{SomDol00})
\begin{equation}
P_{WDM} (k) = T^2_{WDM} (k) P_{CDM} (k),
\end{equation}
\noindent where the WDM transfer function is approximated by
\begin{equation}
T_{WDM} (k) = \exp\left[ -\frac{k\fsl}{2} - \frac{(k\fsl)^2}{2} \right]
\end{equation}
\noindent and $P_{CDM}$ is the CDM power spectrum which is in turn 
approximated by the following formula
(\cite{KH97}):
\vspace{0.2cm}

\noindent $P(k) =$
{\small \[
\frac{Ak}{\left( 1-1.5598k^{1/2}+47.986k+117.77k^{3/2}+321.92k^2 
\right)^{2 \times 0.9303}}
\]}
\begin{equation}
\end{equation}
\noindent This power spectrum was obtained by a direct fit
to the power spectrum estimated using a Boltzmann code and is normalized
to $\sige = 1.0$, close to the cluster abundance and the 4 yr COBE-DMR 
normalization. Here \sige\ is the rms of mass fluctuations estimated 
with the top-hat window of radius $8 \mpch$. 
\begin{planotable}{rccccrrrrr}
\tablecolumns{9}
\tablewidth{0pt}
\tablecaption{Structural Parameters of Host Halos}
\tablehead{\colhead{$\fsl$} & \colhead{Host halo name tag} & \colhead{\vmax} 
& \colhead{$M_{host}$} & \colhead{$N_{satellite}$} & \colhead{$c_{\rm NFW}$}
& \colhead{$e_1$} & \colhead{$e_2$} & \colhead{$T$} \\
(\mpc) &  & (km/s) & ($1 \times 10^{12}$ \msunh) &  &  &  &  & }
\startdata
0.2\phm{100}  &   I\phm{100}  & 258 & 3.3 &  6\phm{100} & 10.8\phm{100} & 0.38 & 0.32 & 0.86 \nl
              &  II\phm{100}  & 246 & 3.1 &  7\phm{100} &  9.4\phm{100} & 0.22 & 0.19 & 0.87 \nl
0.1\phm{100}  &   I\phm{100}  & 270 & 3.4 & 13\phm{100} & 12.0\phm{100} & 0.41 & 0.29 & 0.76 \nl
              &  II\phm{100}  & 255 & 3.2 & 11\phm{100} & 10.5\phm{100} & 0.31 & 0.26 & 0.85 \nl
              & III\phm{100}  & 241 & 2.1 &  4\phm{100} & 10.2\phm{100} & 0.34 & 0.14 & 0.47 \nl
              &  IV\phm{100}  & 196 & 1.7 & 13\phm{100} &  7.6\phm{100} & 0.41 & 0.26 & 0.69 \nl
0.05\phm{100} &   I\phm{100}  & 271 & 3.5 & 15\phm{100} & 12.6\phm{100} & 0.45 & 0.35 & 0.83 \nl
              &  II\phm{100}  & 258 & 3.3 & 22\phm{100} & 11.1\phm{100} & 0.37 & 0.31 & 0.87 \nl
              & III\phm{100}  & 246 & 2.1 & 13\phm{100} & 13.9\phm{100} & 0.34 & 0.18 & 0.58 \nl
0.0\phm{100}  &  II\phm{100}  & 263 & 3.3 & 35\phm{100} & 11.0\phm{100} & 0.36 & 0.34 & 0.96 \nl
\enddata
\end{planotable}

Since our aim is to study the substructure in Milky Way-size 
halos, we simulate WDM models with three different values of \fsl\
namely 0.2, 0.1, and 0.05 Mpc, for 
which, according to eq. (1), $m_W$ is 605, 1017 and 1711 eV, respectively.
Particle masses of this order were recently proposed with the aim to predict
fewer Milky Way satellites than in the CDM scenario (\cite{KL99};
\cite{WC2000}). For a 1 \kev\ particle and for $\omew \sim 0.3$, the rms velocity 
of the particles is $\sim 2 \kms$ at $z = 40$ (\cite{HD2000}). This velocity 
is too small to affect the structure of our simulated host halos. We thus
do not consider in our initial conditions the thermal component 
contribution to the velocities of the particles.

%=====================

\section{Numerical simulations}

%=====================

A set of high-resolution simulations of Milky Way-size halos (defined
as host halos) have been performed using a
variant of the Adaptive Refinement Tree 
(ART) N-body code (\cite{KKK97}).
The ART code achieves high spatial resolution 
by refining the base uniform grid in all
high-density regions with an automated refinement algorithm. 
This version of the ART code (\cite{KKBP00}) has the ability
to handle particles with different mass. This is used
to increase the mass and spatial resolution in few selected
halos. Following, we describe the way in which the simulations 
have been carried out.

First, we set the number of mass levels in the mass hierarchy
to four in all simulations\footnote{The number of mass levels 
was restricted by the amount of memory of the computer where 
the simulations were performed.}. Particles are eight times more massive
when they pass from one level to the next coarser level. 
For our selected number of mass levels, the mass resolution 
on the finest level corresponds to a box of $256^3$ particles.
The realization of the initial spectrum of perturbations
is done in such a way that this number of particles could
be generated in the simulation box.
The size of the simulation box is defined by the requirement of
high mass resolution and by the total number of particles in 
the finest level. We set a box size of $15\ \mpch$ on a side in
all simulations which gives a mass per particle $m_p = 1.66 
\times 10^7\ \msunh$ on the finest level of mass resolution. 
The omission of wavelenghts larger than $15\ \mpch$ may
affect the characteristic mass and spatial distribution of halos,
however, we do not expect it to influence the halo internal structure
(\cite{FWDE88}). A host halo of about $2 \times 10^{12}\ \msunh$ 
resimulated at high resolution will have more
than $10^5$ particles within its virial radius $r_{\rm vir}$, defined
as the radius at which the average halo density is 334 times 
the background density for our selected cosmology, according to 
the spherical top-hat model. Once the
mass hierarchy is fixed, we start by running a low-mass resolution 
(LMR) simulation with $32^3$ particles in a mesh with $256^3$ cells from
which potential host halos are to be identified 
for their future resimulation and analysis.

Second, the Bound Density Maxima (BDM) group finding algorithm 
(\eg\ \cite{Colin99}) is used to locate potential host halos in 
the LMR simulation. The BDM algorithm finds the positions of 
local maxima in the density field smoothed at the scale of interest 
and applies physically motivated criteria to test whether a group 
of particles is a gravitationally bound halo. The friends of 
friends group finding algorithm is then used on these halo 
population to identify the degree 
of isolation of each halo (halos are also visually located).
%ref2
Only halos which are relatively isolated\footnote{There are a 
couple of groups with a mass $\simgreat 10^{13}\ \msunh$ in our
box. Milky Way-size halos belonging to one of these two groups
were rejected for a subsequent analysis.} were
taken as potential host halos because we are interested 
in comparing our numerical results with the observed substructure
in the Milky Way and Andromeda galaxies (\cite{KKVP99}; \cite{Moore99a}).

Third, we trace back all particles within a radius $r\simgreat 1.5\ r_{\rm vir}$
of each selected host halo to get their lagrangian positions at $z = 40$. This
radius is large enough to keep the contamination 
due to the presence of particles from the second mass level at the level
of $ \simless 1$\% in mass. We then regenerate the initial distribution using all 
particles with the four different weights (1, 8, 64, 512 $\times m_p$).
The farther away the particle is from the host halo
the more massive the particle is. The simulation with $\fsl = 0.1\ \mpc$, 
for example, has 1,722,295 particles distributed as follows: 1,450,496 
particles in the first-mass-level, 218,112 in the second-mass-level, and 
31,040 and 22,647 in the third- and fourth-mass levels, respectively. 
The initial conditions are again evolved using ART with the capability of 
handling particles with different mass. The formal force resolution, 
measured by the size of a cell in the finest refinement grid,
is 0.45 \kpch\ and the number of time steps varies from 325 to 41600. 
Accurate results are expected at distances four times larger than
the formal resolution.

Fourth, the BDM is used once again now to identify satellites (guest
halos) orbiting around the center of mass of
host halos. One of the parameters of BDM is the number of spheres that
are randomly placed on the box to locate local maxima ``seeds''. We made sure
not to miss a significant fraction of guest halos 
by using the position of every fourteenth 
first-mass-level particle, a number which is much higher than the expected
number of halos. For example, for the $\fsl = 0.1\ \mpc$ model we used
about $10^5$ seeds. Some guest halos have been polluted by more than 5\% in
mass with particles from the second-mass level. 
These guest halos are usually at the periphery of the host halos and 
thus are more susceptible to being contaminated.
We keep them in our 
satellite catalogs because we think they would still be there even if we 
increased the size of the high-resolution region. To measure the effect on the
number of satellites due to the smallness of the high-resolution volume
an additional simulation for
one of the host halos ---the halo II in the model with $\fsl = 0.1\ \mpc$---
was run doubling the radius of the high-resolution region ($\sim 3\ r_{v}$).
The number of satellites within a sphere of radius
200 \kpch\ centered on the host halo is equal to 11 in both simulations.
This does not mean that there are not any differences at all; for example,
a satellite, which was close to the center ($\sim 30\ \kpch$) 
in the simulation with the smaller high-resolution region,
disappears in the test simulation. However, these differences are
negligible as far as the cumulative circular velocity function for
satellites is concerned.

In Table 1 we present the values of some of the physical 
properties of the host halos re-simulated at
high-resolution for each WDM model (\ie\ for each selected \fsl\ value). 
The name tag of the host halo and its maximum circular velocity 
\begin{equation} 
	V_{\rm max} =\left( \frac{GM(<r)}{r} \right)^{1/2}_{\rm max},
\end{equation}
\noindent where $M(<r)$ is the mass of the halo inside radius $r$,
are placed in the second and third column, respectively. The virial mass
and the number of satellites within a sphere of radius $200 \kpch$,
centered on the host halo, are displayed in the fourth and fifth column, 
respectively. All satellites with more than 10 bound particles 
are counted here. The concentration parameter $c_{\rm NFW}$ 
(\cite{NFW97}) is shown in the sixth column. It is defined here
as the ratio between $r_{\rm vir}$ and $r_s$, where $r_s$ is the NFW
scale radius.
Because our simulations possess the same seed we were able to identify
a couple of host halos (I and II) in our three WDM models and make 
an inter-comparison study of their structural properties. 

We have also measured the ellipticities of the host halos using the tensor of
inertia. This is defined as
\begin{equation}
I_{i,j} = \sum x_i x_j / r^2,
\end{equation}
where the sum is over all DM particles within $r_{vir}$, $x_i$ ($i=1,2,3$)
are the coordinates of the particle with respect to the center of mass
of the halo, and $r$ is the distance of the particle to the halo center.
The ellipticities are then given by
\begin{equation}
e_1 = 1 - \frac{\lambda_1}{\lambda_3},\ \ \ \ \ \ \ \ \ \ \  \ e_2 = 1 - \frac{\lambda_2}{\lambda_3},
\end{equation}
where $\lambda_3 > \lambda_2 > \lambda_1$ are the eigenvalues of the tensor
of inertia. We evaluate the triaxiality parameter using the following
formula (\eg \cite{FIdZ90})
\begin{equation}
T = \frac{\lambda_3^2 - \lambda_2^2}{\lambda_3^2 - \lambda_1^2}
\end{equation}
A value of $T = 1.0$ means that the halo is prolate while a value
of $T = 0.0$ represents an oblate halo. The ellipticities $e_1$, $e_2$,
and the triaxial parameter for host halos are shown in the last three
columns. The ellipticies increase as \fsl\ diminishes. In particular,
the increment is by almost a factor of 
two for the host halo II when one goes from the WDM model with $\fsl = 0.2\ \mpc$ 
to the CDM model. 

Figure 1 provides a visual example of guest halos found in our diffferent
models, including a CDM model. As the host halo we have selected  the
one denoted by the roman number II in Table 1. This halo has a 
$\vmax \approx 250 \kms$ but it varies a little from model to model. 
The letter (a) identifies the halo in the CDM model ($\fsl = 0$) 
and letters from (b) to
(d) represent the halo in WDM models from $\fsl = 0.2\ \mpc$ to 
$\fsl = 0.05\ \mpc$, respectively. It is notable the absence of
substructure and a greater roundness morphology of the host halo 
in our $\fsl = 0.2\ \mpc$ model as compared with the CDM one, which
agrees with the analysis of the host halos ellipticities of
the previous paragraph.

%================

\section{The cumulative circular velocity function of satellites}

%================

The first interesting result that arises from our WDM
simulations is that despite the power spectrum is suppressed 
below the free streaming scale, halos of size close to or smaller 
than \fsl\ are formed. This result 
is not obvious at all. Only high-resolution
numerical simulations could show whether galactic substructures 
would form and survive in such a scenario with a filtered power spectrum at 
high wavenumbers.   
The number of satellites per host halo increases as \fsl\ 
decreases (see Table 1). This is in agreement with the numerical
results at $z = 3$ of \cite{WC2000}.
Their cumulative number of halos above a certain mass increases as 
the cut-off wavenumber $k_0$ increases (the power spectrum suffers 
a sharp drop at $k_0$).

The present-day cumulative maximum circular velocity
satellite functions, $N(> \vmax)$, for our four models are
displayed in Figure 2. These functions were
estimated as follows: for 
each \fsl\ value we count the number of satellites with \vmax\
greater than a given value within 200 \kpch\ from the
center of host halos. This number is then divided by the number
of host halos for each model and the volume of a sphere of 200 \kpch\
radius. Although we plot this function down to $\vmax \sim 10\ \kms$ we
are probably complete only to $\vmax \sim 20\ \kms$ (KKVP). 
As expected, the number of satellites is much smaller than 
the one predicted by a CDM model (KKVP; \cite{Moore99a}; Table 1 this paper).
The WDM model that seems to reproduce better the observed $N(> \vmax)$ function
(taken from KKVP) is that with $\fsl = 0.1\ \mpc$ or 
$m_w \approx 1\ \kev$. The discrepancy seen in Figure 2 between 
simulations and observations for $\vmax \simgreat 50\ \kms$ can be 
attributed to the intrinsic  dispersion  of the mass aggregation 
history of host halos (\cite{BKW2000}).

Is the reduced number of satellites at present time in the WDM scenario
caused only by the suppresion of power at small scales? To answer this question we 
counted the number of guest halos in a sphere with proper radius 
200 \kpch\ centered on
the host halo at $z = 1$, which is close to the
nominal epoch of formation of host halos, and compared it with the
number we have at $z = 0$.
We did this only for the halo II\footnote{
The BDM algorithm finds two progenitors of the host halo I
of comparable mass at $z = 1$, so we decided not to use this halo
for the argument developed in the paragraph.} for which a CDM 
simulation had also been performed. At $z=1$ we obtain 29, 27, 20, 
and 10 guest halos for $\fsl = 0,\ 0.05,\ 0.1,\ 0.2\ \mpc$, respectively
while at $z = 0$ the corresponding numbers are 28, 13, 9, and 4. 
Only halos with \vmax\ greater than 20 \kms\ were chosen and this 
lower limit on \vmax\ was increased by 20\% for halos at $z = 1$ to take
into account the evolutionary reduction of \vmax\ (\eg\ \cite{Colin99}).
%{\pspicture(0.5,-1.5)(12.0,11.3)
{\pspicture(0.5,-3.0)(12.0,11.3)
\rput[tl]{0}(0.8,10.5){\epsfxsize=7.3cm
\epsffile{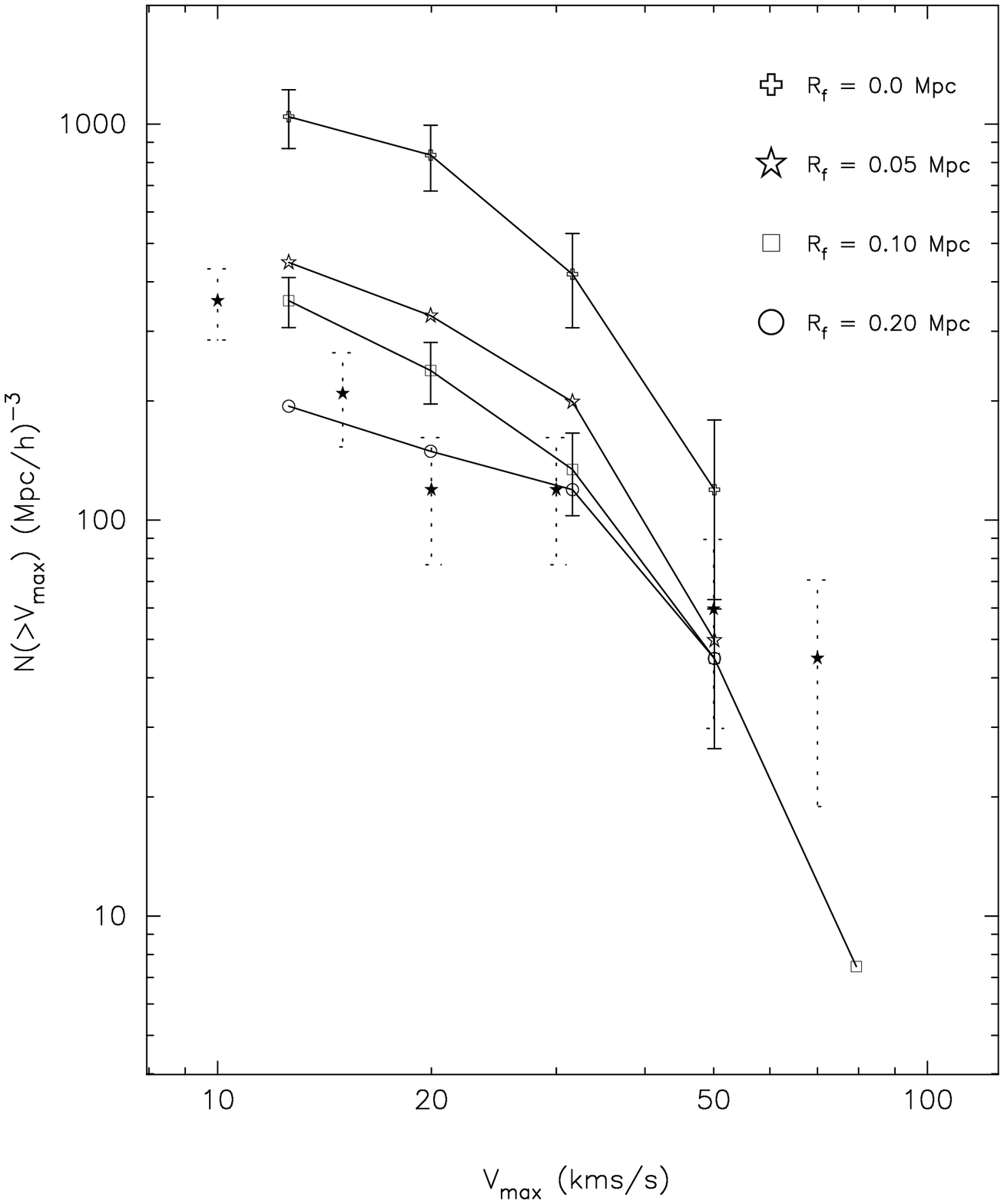}}
\rput[tl]{0}(0.4,1.5){
\begin{minipage}{8.8cm}
  \small\parindent=3.5mm {\sc Fig.}~2.---
The cumulative maximum circular velocity \vmax\ function for
satellites within 200 \kpch\ from the center of the host halo. Solid lines represent
the averaged \vmax\ function for each of our model, from top to bottom as
\fsl\ goes from 0.0 to 0.2 \mpc. The averaged \vmax\ function from satellites 
of Milky Way and Andromeda is represented by stars (taken from KKVP).
Error bars are just Poisson errors. For clarity, we only plot error bars
for the CDM model and the WDM model with $\fsl = 0.1\ \mpc$ and the
observed \vmax\ function is displayed with dotted error bars.
The square at $\vmax \sim 80\ \kms$
has no error bar because there is only one guest halo at this bin.
\end{minipage}}
\endpspicture}

The number of guest halos for the CDM model remains approximately constant 
within this proper volume after $z = 1$ (\cite{Moore99a}). Does
this mean that the guest halos inside this volume are not destroyed and
that small halos outside this volume do not fall later into this volume? 
We have done the following experiment to measure the degree of satellite
destruction (\cite{Kravtsov2000}): 
we have tagged all DM particles within 200 \kpch\ from the
center of the host halo at $z = 0$ and traced back them at $z =1$. We
then apply the BDM algorithm on this subset of DM particles at $z = 1$.
The difference between the number of halos found at $z =1$ and its
number at $z = 0$ gives us
the degree of satellite destruction. The percentage of destruction
is about 35\% for the CDM model while it is 63\% for the WDM model with
$\fsl = 0.1\ \mpc$. Since there is indeed destruction of guest halos 
from $z = 1$ to $z = 0$, the only manner to maintain a constant number
of satellites in the CDM model is by incorporating small halos from outside
the chosen volume (actually, some small halos might also form within the 
volume; however, the probability of such an event is very low).
The accretion rate of guest halos is less in the WDM models
just because the number of available halos that could fall 
into the volume is lower than in the CDM case. We have also 
detected that guest halos are more easily destroyed in the WDM models. These
two effects which are of comparable size work together to deliver a much smaller
number of satellites at $z = 0$.

The more efficient disruption of satellites in the WDM scenario is very
likely due to the fact that guest halos 
are less concentrated in this scenario (see next section).
In fact, we have found that the guest halos that survived until $z=0$ 
are more concentrated than those at $z = 1$. 
On average, the guest halos at $z \approx 1-1.5$  are twice less concentrated 
than those at $z=0$.

%==============================================================
\section{Concentrations and density profiles}
%==============================================================

We define the concentration parameter $c_{1/5}$  as the ratio between
the halo radius $r_h$\footnote{The halo radius $r_h$ is deffined as 
the minimum between $r_{\rm vir}$ and the 
truncation radius (where the spherically averaged outer density profile 
flattens or even increases). In fact, all the host halos and most of the 
guest halos (more than 70\%) attain their virial radius.} and the radius 
within which 1/5 of the total halo mass $M_h$ is contained.   
In Figure 3 we plot this parameter versus the halo 
mass for host (large symbols) and guest (small symbols) halos for
our WDM model with $\fsl = 0.2\ \mpc$ (triangles) and the CDM model
(crosses). Only those halos which 
have more than 90 particles inside their radii have been analyzed. 
The solid and dashed
lines in Figure 3 are extrapolations to small masses of the relations
$c_{1/5}-M_h$ found by Avila-Reese et al. (1999) for the corresponding
CDM model for isolated halos and halos in groups, respectively.
From Figure 3 one can see that the concentration 
of host WDM halos is only slightly smaller than that of
CDM halos. For the small guest halos, the difference
is more notable; in the $10^9-10^{11}\ \msunh$ mass
range the concentrations for the case $\fsl = 0.2\ \mpc$ are approximately 
$1.8-1.2$ times smaller than those obtained in the CDM model.
It is interesting to see that the extrapolations showed in Figure 3
do actually agree with the concentrations of the guest halos in
our fiducial CDM model.

The density profiles of host halos obtained in our 
simulations with $\fsl = 0.2$, 0.1, 0.05, and 0.0\ \mpc\
($M_h \approx 1-3 \times 10^{12}\ \msunh$) are shown
in Figure 4 (symbols). In order to avoid overlapping, the 
profiles from the $\fsl = 0.1$, 0.05, and 0.0 \mpc\ models were 
shifted in $\log \rho$ by $-1$, $-2$, and $-3$, respectively. 
These density profiles are well described by the NFW formula 
(lines), although the inner slope in some cases of the WDM models
is slightly shallower than $r^{-1}$. 
The inner profile for the CDM halo is actually 
steeper than $r^{-1}$. The corresponding NFW concentration
parameters for the host halos are given in Table 1. 
For comparision, in the CDM simulation of \cite{AFKK99} a $10^{12}\ \msunh$ 
halo has in average $c_{\rm NFW} \approx 12$.  
 
The guest halos ($M_h \approx 10^9-10^{11}\ \msunh$) present a wide diversity of 
density profiles. Unfortunately, since the number of particles in these halos
is not very large ($\sim 100-200$ particles for most of them), the 
resolution is not sufficient to study the inner density profile with
accuracy (see a discussion on this subject in $\S 6.2$). For those 
guest halos ($\sim 15\% $) whose density profiles are reasonably 
well described by the NFW fit we obtain a mean $c_{\rm NFW} \approx 8$ 
and 10 for $\fsl = 0.2$ and 0.05 \mpc\ models, respectively. 
We estimate a mean $c_{\rm NFW}$ of 25 by extrapolating
the results of the CDM model to low masses (\cite{AFKK99})
or by using our own CDM simulation.
{\pspicture(0.5,-1.5)(12.0,11.3)
\rput[tl]{0}(1.0,10.5){\epsfxsize=8cm
\epsffile{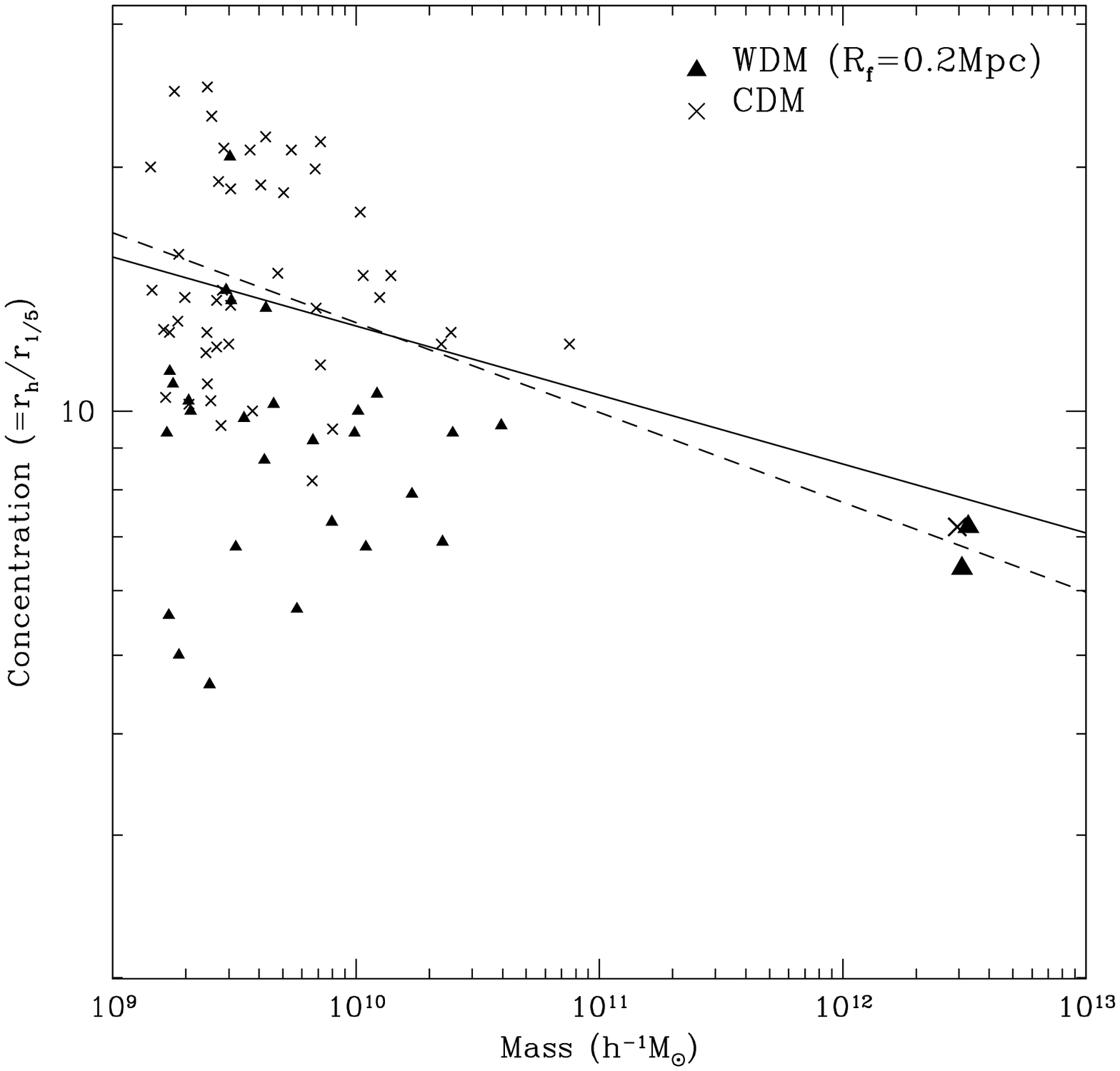}}
\rput[tl]{0}(0.4,2.0){
\begin{minipage}{8.8cm}
  \small\parindent=3.5mm {\sc Fig.}~3.---
The concentration parameter $c_{1/5}$ versus the mass
for host halos (large symbols) and guest halos with more than
90 particles (small symbols) in the WDM ($R_{f,WDM}$ = 0.2, 
solid triangles) and CDM (crosses) simulations. The 
solid and dashed lines correspond to the 
linear fittings found in a CDM simulation for isolated
halos and halos within group- and galaxy-size halos, respectively
(\cite{AFKK99}). The only host CDM halo here resulted less
concentrated than the average.
\end{minipage}}
\endpspicture}

\section{Discussion}

\subsection{The accretion and destruction rates of guest halos}

Our results show that in a WDM model with $\fsl\ = 0.1\ \mpc$ the 
number of guest halos within Milky Way-size halos
agrees with the observed number of satellites in the Milky Way
and Andromeda galaxies. 
Using the extended Press-Schechter formalism and assuming 
that {\it the amount of substructure is preserved since the host 
halo formation epoch}, Kamionkowski \& Liddle (1999) also found good 
agreement with the observations for models with a sharp drop 
in the power spectrum on small scales and with zero velocity 
dispersion (they were interested in studying the 
broken-scale-invariant model). However, the cut-off length-scale 
in the power spectrum they requiered is two to three times smaller 
than the one used here in order to fit the observational data.
We argue that at the basis of this difference is the fact that
for models with a cut-off in the power spectrum and with 
negligible particle dispersion velocities,
the amount of substructure in a host halo {\it is not} preserved.

{\pspicture(0.5,-1.5)(12.0,11.3)
\rput[tl]{0}(0.5,10.5){\epsfxsize=8cm
\epsffile{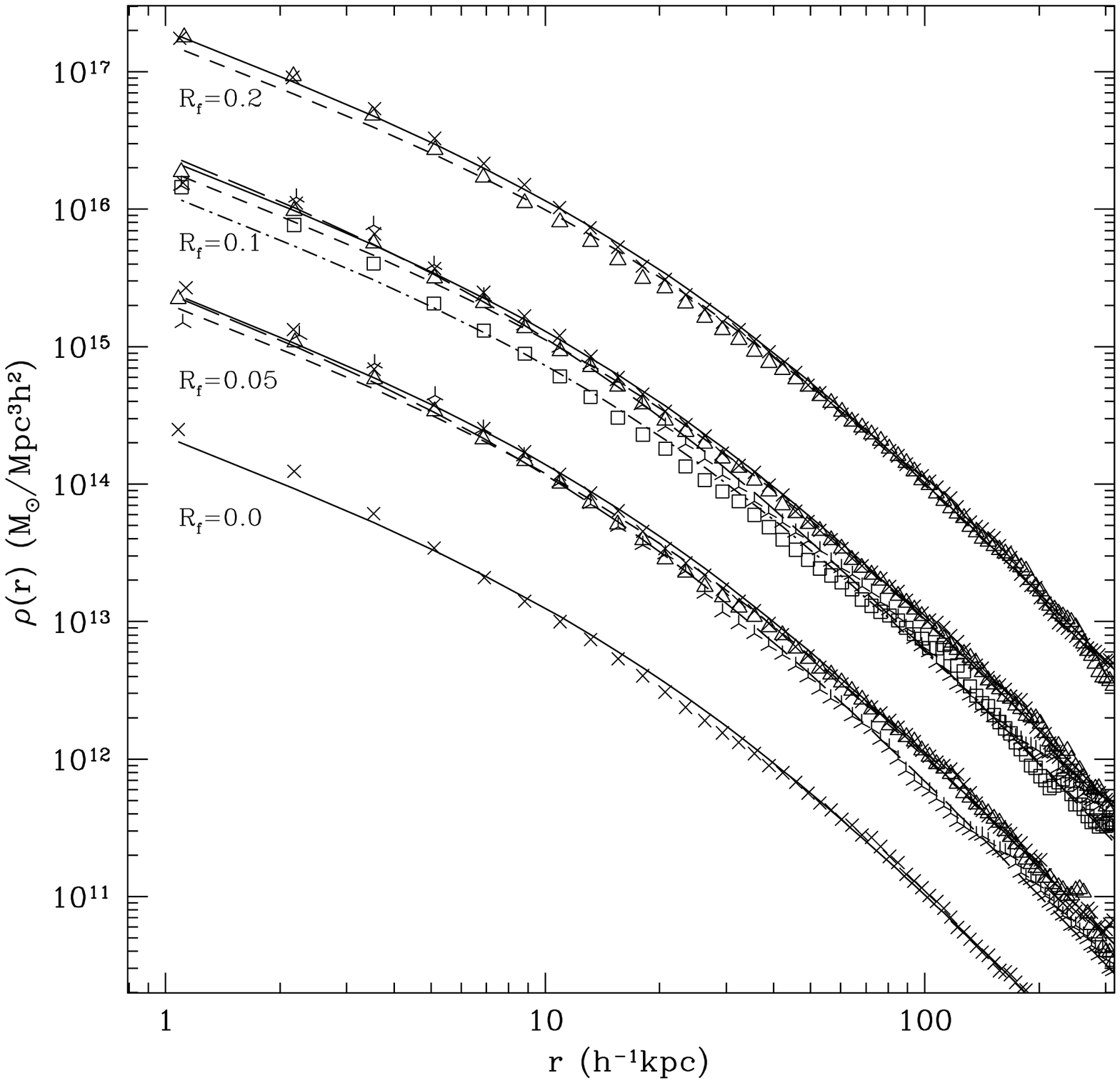}}
\rput[tl]{0}(0.0,2.0){
\begin{minipage}{8.8cm}
  \small\parindent=3.5mm {\sc Fig.}~4
The density profiles of the host halos in the
WDM models with $R_{f,WDM}$ = 0.2, 0.1 and 0.05 \mpc, and in 
the CDM model ($R_{f,WDM}$ = 0.0).
The profiles of the three latter simulations were shifted in the
log of the density by $-1$, $-2$, and $-3$ in order to avoid overlapping.
The different lines are fits to the data using the NFW profile and
have the same correspondence with the symbols in each one of the
models.
The corresponding NFW concentration parameters are given in Table 1.
\end{minipage}}
\endpspicture}
  
In Section 4, we have shown that for the CDM model the number 
of satellites within a sphere of proper radius 200 \kpch\ centered
on the host halo remains approximately constant from the nominal 
epoch of host halo formation up to the present time (see also 
\cite{Moore99a}). Nevertheless, as discussed in $\S~4$, this is 
a particular case in which the destruction and accretion/formation 
rates of guest halos balance up. In WDM models,
the accretion/formation rate of guest halos is smaller and the destruction rate 
is more efficient than the one in the CDM scenario.
These two facts operate at the same time to produce
a number of guest halos at $z = 0$ much smaller than 
the one found at the nominal epoch of host halo formation.  
Therefore, Kamionkowski \& Liddle (1999) overestimated the filtering scale of 
the power spectrum because they did not consider that a high fraction of 
satellites are destroyed by dynamical effects during their
lifetime in the dense environment of host halos. Recently, White \& 
Croft (2000) have reported from numerical simulations of WDM
models that at $z = 3$ the number of small halos is reduced 
roughly by a factor of five when the CDM spectrum is filtered at 
scales much larger than 0.1\ \mpc. If one takes into account 
that a significant fraction of these halos will be destroyed by $ z = 0$
and not many small halos will be accreted,
then the agreement with observations will be reached for  
a cut-off scale smaller than the one suggested by these authors.

Why a significant fraction of guest halos become destroyed in the WDM scenario?
We offer the following answer: in a WDM model these halos form later
than in a CDM model, when the mean density of the universe was lower. For
example, the formation redshift (defined as the redshift 
where $\sigma (M_{\rm vir},z) = 1$)
of a halo of $M_{\rm vir} = 2 \times 10^9 \msunh$ is 4.1 for
the CDM model, whereas for the WDM model with $\fsl = 0.2\ \mpc$ it is 2.5.
Thus, the characteristic overdensity $\delta_c$ for guest halos 
of this mass in this WDM model is about 4.5 lower than the 
corresponding for guest halos in the CDM model. If $\delta_c$ 
approximately scales as the cube of the concentration parameter, this 
would mean that these halos were about 1.7 less concentrated
at their formation epoch. This interpretation agree with the results from
our study on concentration parameters ($\S 5$). In summary, guest 
halos in WDM models are more easily disrupted because of their 
puffier density distribution
 
\subsection{Density profiles of the WDM halos}
We have found that the present-day density profiles of Milky Way- 
size galactic halos in a WDM model with a \fsl\ value
corresponding to masses of the order of a few $10^9\ \msun$
do not differ from those obtained in CDM models. A similar result
was also reported by Moore et al. (1999b). They cut the power spectrum
on a mass scale smaller by more than a factor of ten than the 
mass of the studied cluster. The question about how the 
filtering of the power spectrum affects the inner density profiles 
of halos with masses close or smaller than the cut-off mass scale
is still open (see Avila-Reese et al. 1998). 
Unfortunately, we could not explore this
question in detail in this paper because our guest halos with masses
close to $2 \times 10^9\ \msun$ are poorly resolved (a 
study on this direction is in progress; small halos in a WDM
cosmology are being resimulated with high-resolution). Nevertheless, 
we have found that the small WDM 
halos are actually less concentrated than the CDM ones.
The difference in the mean in our extreme $\fsl = 0.2\ \mpc$ model
is less than a factor of two in the $c_{1/5}$ parameter. We 
estimate that this factor is too small to offer 
a solution to the problem related to the inner structure
of dwarf galaxy dark halos. Thus, although the WDM scenario helps, 
apparently it does not solve this problem. Nevertheless, some questions
that remain open might change this conclusion. In the following, we 
discuss them.

 In the monolithic collapse the orbital tangential velocity of the 
collapsing particles plays a significant role in the final virialized 
configuration; it is expected that in a hot monolithic collapse 
shallow cores form (\eg\ \cite{vanAlbada82}; \cite{AM90}; \cite{Firmani2000b}).
To produce a hot monolithic collapse in a cosmological scenario,
dark matter particles should have a non-negligible velocity 
dispersion (thermal energy) at the redshift of the halo formation
--- there should also be a cut-off in the power spectrum at the
scale of interest.
On the other hand, if this is the case, then the formation of shallow
cores with a maximum limiting density determined by the velocity
dispersion is expected (\cite{HD2000}). These authors 
estimate that shallow cores, which are in agreement 
with observational determinations in dwarf galaxies, can be produced 
if warmons have a mass $m_W \approx 200\ \ev$; in this case the 
particle dispersion velocity is significant. Of course, according to 
our results, a WDM model with $m_W \approx 200\ \ev$ would produce too
few guest halos to agree with observations. However, if the inclusion
of gas within the guest halos plays an important role to avoid
their disruption, then a larger fraction of guest halos could survive
until the present epoch. It is probable that with an $m_W$ only slightly
smaller than 1\kev, shallow cores that explain the rotation curves
of dwarf galaxies form if the velocity dispersion is taken into 
account. Nevertheless, as we have shown, in this case shallow cores 
in the massive (host) halos will not be produced. Thus, if the increment 
of the core radius with the halo mass or \vmax\ inferred by Avila-Reese 
et al. (1998) and Firmani et al. (2000b) from observations is confirmed, 
then the collisionless WDM scenario is ruled out.
It is interesting to note that in a self-interacting WDM model,
the particle velocity dispersion would be larger than in 
the collisionless case (\cite{Hannestad2000}). A detailed numerical
study of this model, which appears to be more
palatable than a self-interacting CDM model, is desirable.

Finally, we should comment that the problem of an apparently excessive
number of guest halos in the CDM model holds if one assumes
that each guest halo hosts a luminous galaxy. This assumption was 
recently challenged by Bullock et al. (2000) who proposed that 
reionization can efficiently inhibit dwarf galaxy formation. 

\section{Conclusions}

Using high-resolution N-body simulations, we have studied the substructure 
inside Milky Way-size halos in a WDM cosmological scenario. 
We have also addressed the question whether the
density profiles of host and guest halos are different from their
corresponding CDM ones. Our main conclusions are:
 
1. Despite the fact that the power spectrum of fluctuations is suppressed at
small scales, a non-negligible number of virialized structures,
corresponding to these or smaller scales, 
form and survive within larger structures.
The accretion rate of small halos is found to be less in the
WDM scenario than in the CDM one. This is simply explained by the fact
that a smaller number of small halos are avalaible for their incorporation into
host halos in the WDM models. A higher satellite destruction rate is found in the
WDM scenario as compared with the one in the CDM model: it
can be accounted
for the fact that guest halos are less concentrated by about a factor two
in average in the WDM models. The less efficient halo accretion and the higher 
satellite destruction have almost the same weight
as far as the final count of satellites within host halos at $z = 0$ is concerned.
The larger the \fsl\ value the greater the size of these two effects,
and the smaller the abundance of satellites.

2. The predicted maximum circular velocity 
function of guest halos that seems to best fit the observed one for
satellites in the Milky Way and Andromeda
is that given by the $\fsl=0.1\ \mpc$ model. This \fsl\ value
corresponds to a warmon of mass about 1 keV.

3.  For the $\fsl =0.1$ and 0.2\ \mpc\ models, guest halos 
($M_h \approx 10^9-10^{11}\ \msunh$) have 
a concentration parameter $c_{1/5}$ which is roughly 
twice smaller than that of the CDM halos.  
For those guest halos whose density profiles are reasonably
well described by a NFW parametric fit ($\sim 15\%$),
the $c_{\rm NFW}$ parameter is
roughly $1.5-3.0$ times lower than that of 
the CDM halos. This difference in the concentration parameters, 
for both $c_{1/5}$ and $c_{\rm NFW}$, vanishes as we go to 
more massive halos.
  
4. The density profile of the host halos ($M_{\rm vir} \approx 
1-3 \times 10^{12}\ \msunh$) is well described by the
NFW profile (in some cases, the inner slope is slightly shallower
than $r^{-1}$). The guest halos have a wide variety of density
profiles and those whose masses are below the corresponding
cut-off mass scale probably present a small shallow 
core. The poor mass resolution of the simulations at these 
scales limits our predictions. 
 
In summary, we have shown that in the WDM model with 
$\fsl \approx 0.1 \mpc$ or $m_W \approx 1\ \kev$ 
the degree of substructure within a Milky Way-size
halo is much lower than in the CDM model and is in agreement 
with observations, if one assumes that each guest halo hosts a luminous galaxy.
The problem of cuspy halos probably still persists in
the WDM scenario, although we have found that halos ---in particular the
small ones--- are less concentrated than the corresponding
CDM halos. If the inclusion 
of baryonic matter helps to significantly avoid disruption of substructure,
then the agreement with observation may continue
for less massive WDM candidates for which the rms velocity is 
larger. In this case, the formation of shallow cores in dwarf galaxy like halos
is expected. Nevertheless, as we have shown, halos of larger masses 
will not have a shallow core.

\acknowledgments
 
We are grateful to Anatoly  Klypin and Andrey Kravtsov for kindly 
providing us a copy of the ART code in its version of multiple mass, and for
enlightening discussions. We also acknowledge the anonymous referee
whose helpful comments and suggestions improved some aspects of 
this paper. P. C. and V.A. acknowledge the hospitality of the NMSU 
Astronomy Department. Our ART simulations were performed at the 
Direcci\'on General de Servicios de C\'omputo Acad\'emico, UNAM, using 
an Origin-2000 computer. 

%=======================
% References
%=======================

\end{document}